\documentstyle[prd,aps,preprint,tighten,floats]{revtex}
\begin{document}
\draft
\preprint{UPR-649-T}
\date{March 1995}
\title{All the Four Dimensional Static, Spherically Symmetric
Solutions of  Abelian Kaluza-Klein Theory}
\author{Mirjam Cveti\v c
\thanks{E-mail address: cvetic@cvetic.hep.upenn.edu}
and Donam Youm\thanks{E-mail address: youm@cvetic.hep.upenn.edu}}
\address {Physics Department \\
          University of Pennsylvania, Philadelphia PA 19104-6396}
\maketitle
\begin{abstract}
{We present the explicit form for all the four dimensional,
static, spherically symmetric solutions in $(4+n)$-d Abelian
Kaluza-Klein  theory by performing a subset of $SO(2,n)$
transformations corresponding to four $SO(1,1)$ boosts on the
Schwarzschild solution, supplemented by $SO(n)/SO(n-2)$ transformations.
The solutions are parameterized by the mass $M$, Taub-Nut charge $a$,
$n$ electric $\vec{\cal Q}$ and $n$ magnetic $\vec{\cal P}$ charges.
Non-extreme black holes (with zero Taub-NUT charge) have either the
Reissner-Nordstr\" om or Schwarzschild global space-time. Supersymmetric
extreme black holes have a null or naked singularity, while
non-supersymmetric extreme ones have a global space-time of extreme
Reissner-Nordstr\" om black holes.}
\end{abstract}
\pacs{04.50.+h,04.20.Jb,04.70.Bw,11.25.Mj}

Theories that attempt to unify gravity with other forces of nature in
general involve, along with the graviton, additional scalar
fields.  Non-trivial 4-dimensional (4-d) configurations for such
theories include a spatial variation of scalar fields, which in turn
affects the space-time and thermal properties of such configurations.
In particular, spherically symmetric solutions in Einstein-Maxwell-dilaton
gravity have been studied extensively \cite{HOROWITZ}.
A subset of such configurations corresponds to black holes (BH's)
which arise within effective (super)gravity theories describing
superstring vacua.  Configurations arising in the compactification of
$(4+n)$-d gravity, {\it i.e.}, Kaluza-Klein (KK) theories \cite{KAL},
are also of interest since KK theory attempts to unify gravity with
gauge interactions.  In addition, such configurations can be viewed as
a subset of BH's within the effective 4-d theory of heterotic superstring
vacua\cite{HET,DUFF}.

In this letter, we find the explicit form for all the static, spherically
symmetric solutions in $(4+n)$-d Abelian KK theory.  These results as well
as analogous results for BH's in effective string theory \cite{GKS}
were anticipated in Ref. \cite{BMG}, where the existence of a general
class of solutions, which are obtained by appropriate generating
techniques, was proven, however, without explicit calculations of the
sort we shall present here.  Such solutions can be generated by a subset
of the $SO(2,n)$ ($\subset SL(2+n, \Re)$) transformations on the
Schwarzschild solution.  The explicit form of the 4-d space-time metric
allows for the study of the global space-time and the thermal properties of
such configurations.  The study generalizes previous studies
\cite{DM,FIV,GW} of BH's in 5-d KK theory, as well as recent studies
\cite{SUPER,WARSAW,NONEX} of BH's with constrained charges in $(4+n)$-d
Abelian KK theory.
In addition, the work sets a stage for generating general axisymmetric
solutions in KK theory \cite{CYV} as well as in other sectors of
supergravity theories \cite{CYVI}.

The starting point is the effective 4-d Abelian KK theory obtained from
$(4+n)$-d pure gravity by compactifying the extra $n$ spatial coordinates
on a torus by using the following KK  metric Ansatz:
\begin{equation}
g^{(4+n)}_{\Lambda \Pi} \equiv
\left [ \matrix{{\rm e}^{-{1 \over \alpha}\varphi}g_{\mu\nu} +
{\rm e}^{{2\varphi} \over {n\alpha}}\rho_{ij}A^i_{\mu} A^j_{\nu} &
{\rm e}^{{2\varphi} \over {n\alpha}}\rho_{ij}A^j_{\lambda}
\cr {\rm e}^{{2 \varphi} \over {n\alpha}}\rho_{ij}A^i_{\pi} &
{\rm e}^{{2\varphi} \over{n\alpha}}\rho_{ij}} \right ],
\label{ansatz}
\end{equation}
where $g_{\mu\nu}$ is the 4-d Einstein frame metric
\footnote{The convention for the signature of metric in this paper is
$(+++-)$ with the time coordinate in the fourth component.}
$A^i_\mu$  are $n$  $U(1)$ gauge fields,  $\rho_{ij}$ is the
unimodular part of the internal metric $g^{(4+n)}_{i+4,j+4}$ and
$\alpha = [(n+2)/n]^{1/2}$.

Static or stationary solutions are invariant under the time-translation,
which can be considered along with $n$ internal $U(1)$ gauge
transformations as a part of $(n+1)$-parameter Abelian isometry
group generated by the commuting Killing vector fields
$\xi^{\Lambda}_i := \delta^{i+3}_i$  ($i=1,\cdots ,n+1$)
of a $(4+n)$-d space-time manifold $M$.  In this case, the projection
of the $(4+n)$-d manifold $M$ onto the set $S$ of the orbits of the
isometry group in $M$ allows one to express the $(4+n)$-d
Einstein-Poincar\'e gravity action as the following effective 3-d one
\cite{M,DM}:
\begin{equation}
{\cal L}= -{1 \over 2}\sqrt{-h}[{\cal R}^{(h)} -
{1\over 4}{\rm Tr}(\chi^{-1}\partial_a \chi \chi^{-1} \partial^{a}\chi)],
\label{threelag}
\end{equation}
where $h_{ab} \equiv \tau g^{\perp}_{ab}$ ($a,b=1,2,3$) is the rescaled
metric on $S$ and
\begin{equation}
\chi \equiv \left [ \matrix{\tau^{-1} & -\tau^{-1}\omega^{T} \cr
-\tau^{-1}\omega & \breve{\lambda} + \tau^{-1} \omega \omega^{T}}\right ]
\label{scalar}
\end{equation}
is the $(n+2) \times (n+2)$ symmetric, unimodular matrix of scalar
fields on $S$.   Here, $\breve{\lambda}_{ij} \equiv g^{(4+n)}_{\Lambda\Pi}
\xi^{\Lambda}_i \xi^{\Pi}_j$, $\tau \equiv {\rm det}\breve{\lambda}_{ij}$
and $g^{\perp}_{ab} \equiv g^{(4+n)}_{ab} - \breve{\lambda}^{ij}
\xi_{ia}\xi_{jb}$.  The ``potential'' $\omega^{T} \equiv
(\omega_1,...,\omega_{n+1})$ defined as $\partial_a \omega_i =
\omega_{ia} \equiv \hat{\epsilon}_{abc}\xi^{b;c}_i$ ($\hat{\epsilon}_{abc}
\equiv \epsilon_{abc4...(4+n)}$) replaces the degrees of freedom of
$\xi_{ia}=g^{(4+n})_{i+3,a}$. The effective 3-d Lagrangian density
(\ref{threelag}) is invariant under the global $SL(2+n, \Re)$ target space
transformations\cite{M}:
\begin{equation}
\chi \rightarrow {\cal U}\chi {\cal U}^{T},\ \ \ \ \ \
h_{ab} \rightarrow h_{ab},
\label{sltran}
\end{equation}
where ${\cal U} \in SL(2+n, \Re)$.  In particular, the $SO(n)$
transformations \cite{NONEX} of the effective 4-d Lagrangian density
constitute a subset of the $SL(2+n,\Re)$ transformations, which
do not affect the 4-d space-time part of the metric.

The physically interesting solutions correspond to the configurations
with an asymptotically ($|\vec{r}| \rightarrow \infty$) flat
4-d space-time metric and constant values of the other 4-d fields.
Without loss of generality one can take the Ansatz:
\begin{equation}
(g_{\mu\nu})_{\infty} = \eta_{\mu\nu},\
(A^i_{\mu})_{\infty} = 0 ,\
\varphi_{\infty} = 0 , \
(\rho_{ij})_{\infty} = \delta_{ij},
\label{asympt}
\end{equation}
which yields $\chi = {\rm diag}(-1, -1,1,...,1)$.

The only subset of $SL(2+n,\Re )$ transformations (\ref{sltran}),
which preserves the asymptotic boundary conditions (\ref{asympt}),
is the $SO(2,n)$  transformation.  A subset of $SO(2,n)$
transformations can then be used to act on known solutions
to generate a new set of solutions of the equations of motion
for the effective 3-d Lagrangian density (\ref{threelag}).

In the following, we shall concentrate on static, spherically
symmetric solutions.  Spherical symmetry implies that for such
configurations the metric $h_{ab}$, in polar coordinates
$(r,\theta ,\phi)$, takes the form:
\begin{equation}
h_{ab} = {\rm diag}\left(1,f(r),f(r)\sin^2 \theta\right),
\label{threemet}
\end{equation}
where $a,b = r,\theta,\phi$, and $\chi$ depends only on the radial
coordinate $r$.  The transformation between the 3-d fields
($h_{ab}$ and $\chi$) and the corresponding 4-d fields is of the form:
\begin{eqnarray}
{\rm e}^{-{\varphi \over \alpha}}g_{\mu\nu}&=&
{\rm diag}(-\tau^{-1}, -\tau^{-1}f, -\tau^{-1}f \sin^2 \theta,
(\breve{\lambda}^{11})^{-1} ),
\nonumber \\
{\rm e}^{{2\varphi} \over {n\alpha}} \rho_{ij}&=&
\breve{\lambda}_{i+1, j+1},\ \  A^i_t = -\breve{\lambda}^{i+1,1}/
\breve{\lambda}^{11},
\nonumber \\
A^i_{\phi}&=& \tau^{-1} f \cos \theta {\rm e}^{{2\varphi}
\over {n\alpha}} \rho^{ij}\partial_r\omega_{j+1} ,
\label{rels}
\end{eqnarray}
with the constraint $\breve{\lambda}^{1k}\partial_r\omega_k = 0$ that
the unphysical Taub-NUT charge is absent.  Here, the spherically symmetric
Ansatz for the 4-d metric is given by $g_{\mu\nu}={\rm diag}
(1/\lambda(r),R(r),R(r){\rm sin}^2 \theta,-\lambda(r))$, and the
4-d scalar fields $\varphi$ and $\rho_{ij}$ depend only on the radial
coordinate $r$.

One way to generate the most general static, spherically symmetric
solutions (with the Ans\" atze (\ref{rels})) is by performing a subset
of $SO(2,n)$ transformations on the 4-d Schwarzschild solution with the
ADM mass $m$, which in terms of the 3-d quantities is of the following form:
\begin{equation}
\chi = {\rm diag}\left(-(1-{m \over r})^{-1}, -(1-{m\over r}),
1,...,1\right),
\label{schthree}
\end{equation}
and $f(r) = r(r-m)$.
The subset of $SO(2,n)$ transformations that generates new types of
solutions is the quotient space $SO(2,n)/SO(n)$
\footnote{All the axisymmetric stationary solutions can be generated
by performing $SO(2,n)/SO(n)$ transformations on the Kerr solution
\cite{CYV}.}.
The $2n+1$ parameters of $SO(2,n)/SO(n)$ along with the parameter
$m$ constitute the $2n+2$ parameters, which correspond to the mass
$M$, $n$ electric $\vec{\cal Q}$ and $n$ magnetic $\vec{\cal P}$
charges as well as the Taub-Nut charge $a$ of the most general,
spherically symmetric, stationary solution in $(4+n)$-d KK theory.
In fact, each representative of the elements of $SO(2,n)/SO(n)$
generates a physical parameter of the solution
\footnote{Similar observations are due to  Gibbons \cite{Gibbons}.}:
$n$ boosts on the first [or the second] index of $\chi$
(of the Schwarzschild solution) and on one of the last $n$ indices of
$\chi$ generate magnetic [or electric] charges, and an $SO(2)$ rotation
on the first two indices of $\chi$ generates an unphysical
Taub-NUT charge $a$.

For the purpose of obtaining the explicit form of static, spherically
symmetric solutions with a general charge configuration, it is
convenient to first perform two successive $SO(1,1)$ boosts on the
$1st$ and $(n+1)$-$th$, and the $2nd$ and $(n+2)$-$th$ indices of
(\ref{schthree}) with the boost parameters  $\delta_{P,Q}$,
respectively, yielding:
\begin{equation}
\chi = \left [ \matrix{-{{r+\hat{P}}\over r} & 0 &\cdot&
{|P| \over r} & 0 \cr
0 & -{{r+2\beta - \hat{Q}} \over {r + 2\beta}} & \cdot&0 &
{|Q| \over {r+2\beta}} \cr
{\cdot}&{\cdot}&{\bf I}& {\cdot}&{\cdot}\cr
{|P| \over r} & 0 & \cdot&
{{r +2\beta - \hat{P}} \over r} & 0 \cr
0 & {|Q| \over {r + 2\beta}} &\cdot&
0 & {{r + \hat{Q}} \over {r + 2\beta}}} \right ],
\label{emchi}
\end{equation}
and $f(r)=r(r -2\beta)$.  Here, $\beta \equiv m/2$ and $\hat{Q} = \beta
+ \sqrt{Q^2 + \beta^2}$  [$\hat{P} = \beta + \sqrt{P^2 + \beta^2}$], where
$P \equiv m{\rm sinh}\delta_P {\rm cosh}\delta_P$ [$Q \equiv m{\rm sinh}
\delta_Q {\rm cosh}\delta_Q$].  ${\bf I}$ is the $(n-2)\times (n-2)$
identity matrix, $\cdot$ denotes the zero entries, and the event
horizon $r_+$ is shifted to the origin ($r=0$).
The solution (\ref{emchi}) corresponds to the $U(1)_M\times U(1)_E$
BH solutions
\footnote{These solutions were first found in Refs. \cite{WARSAW,NONEX}
by directly solving the equations of motion with a diagonal internal
metric Ansatz.}
with the ADM mass $M={\hat P} +{\hat Q}$, the physical magnetic
[electric] charge $P$ [$Q$], and $\beta \ge 0$ measuring a deviation
from the supersymmetric limit \cite{SUPER}.

A class of new solutions can be obtained by performing $SO(n)/SO(n-2)$
transformations, parameterized by $2n-3$ parameters, on (\ref{emchi}).
Such transformations act on the lower-right $n\times n$ part of $\chi$
and, thus, do not affect the 4-d space-time metric $g_{\mu\nu}$ and
the dilaton $\varphi$.  The transformed solutions have $n$ electric
${\vec{\cal Q}}$ and $n$ magnetic  ${\vec{\cal P}}$ charges, subject
to one constraint $\vec{\cal P}\cdot \vec{\cal Q}=0$.

Thus, in order to generate the most general, static, spherically symmetric
solution one needs only one more parameter, associated with $SO(2,n)/SO(n)$
transformations. Such a parameter is provided by two $SO(1,1)$ boosts on
the $1st$ and $(n+2)$-$th$, and the $2nd$ and $(n+1)$-$th$ indices of
$\chi$ in (\ref{emchi}), whose respective boost parameters $\delta_1$
and $\delta_2$ have to be related to one another in order to yield
solutions with no Taub-NUT charge.  The transformed solutions are
of the form:
\begin{eqnarray}
\lambda&=&{{r(r+2\beta)} \over {(XY)^{1/2}}} ,\ \
R=(XY)^{1/2} ,\ \  e^{{2\varphi} \over \alpha}=
{X \over Y} ,
\nonumber \\
\rho_{ij}&=&\delta_{ij}{\rm e}^{-{{2\varphi}\over{n\alpha}}} , \ \
\rho_{n-1,n-1}={{We^{{{2(n-2)}\over {n\alpha}}\varphi}} \over
{(XY)^{1/2}}},
\nonumber \\
\rho_{n-1,n}&=&{{Ze^{{{2(n-2)}\over{n\alpha}}\varphi}} \over
{(XY)^{1/2}}} ,\
\rho_{n,n} = {{(r + \hat{Q})(r + \hat{P})} \over {(XY)^{1/2}}}
e^{{{2(n-2)}\over {n\alpha}}\varphi} ,
\label{gensol}
\end{eqnarray}
where
\begin{eqnarray}
X &=& r^2 + [(2\beta - \hat{P} + \hat{Q})
{\rm cosh}^2 \delta_2 + \hat{P}]r  + 2\beta \hat{Q}
{\rm cosh}^2 \delta_2,
\nonumber \\
Y &=& r^2 + [(2\beta + \hat{P} - \hat{Q})
{\rm cosh}^2 \delta_1 + \hat{Q}]r  + 2\beta \hat{P}
{\rm cosh}^2 \delta_1,
\nonumber \\
W &=& r^2 + [(2\beta + \hat{P} - \hat{Q}){\rm cosh}^2 \delta_1
+(2\beta - \hat{P} + \hat{Q}){\rm cosh}^2 \delta_2]r
\nonumber \\
& &+2[\beta(2\beta - \hat{P} - \hat{Q}) +\hat{P}\hat{Q}]
{\rm cosh}^2\delta_1 {\rm cosh}^2 \delta_2
\nonumber \\
& &+(2\beta - \hat{Q})\hat{P}{\rm cosh}^2 \delta_1 +
(2\beta - \hat{P})\hat{Q}{\rm cosh}^2 \delta_2
\nonumber \\
& &+|P||Q|{\rm cosh} \delta_1 {\rm cosh} \delta_2 {\rm sinh}
\delta_1 {\rm sinh} \delta_2,
\nonumber \\
Z &=& [|P| {\rm sinh} \delta_1 {\rm cosh} \delta_2
+ |Q|{\rm sinh} \delta_2 {\rm cosh} \delta_1 ]r \nonumber \\
& &+ |P|\hat{Q} {\rm sinh} \delta_1 + \hat{P}|Q| {\rm sinh} \delta_2 ,
\label{wxyz}
\end{eqnarray}
with the non-zero electric and magnetic charges and the ADM mass given
by
\begin{eqnarray}
P_{n-1} &=& |P|{\rm cosh} \delta_1 {\rm cosh} \delta_2
+ |Q| {\rm sinh} \delta_1 {\rm sinh} \delta_2 ,\nonumber \\
P_n &=& -(\hat{P} - \hat{Q} +2\beta){\rm cosh} \delta_1 {\rm sinh}
\delta_1 ,
\nonumber \\
Q_{n-1} &=& -(\hat{P} - \hat{Q} - 2\beta){\rm cosh} \delta_2
{\rm sinh} \delta_2 ,\nonumber \\
Q_n &=& |Q|{\rm cosh} \delta_1 {\rm cosh} \delta_2
+ |P|{\rm sinh} \delta_1 {\rm sinh} \delta_2 ,
\nonumber \\
M &=& (2\beta + \hat{P} -\hat{Q}){\rm cosh}^2 \delta_1
+ (2\beta + \hat{Q} - \hat{P}){\rm cosh}^2 \delta_2 \nonumber \\
& &+ \hat{P} + \hat{Q} -4\beta .
\label{genpar}
\end{eqnarray}
Here, the electric fields are given by $E_i =R^{-1}
{\rm e}^{-\alpha\varphi}\rho^{ij} Q_j$ ($i=1,...,n$).
The requirement $\breve{\lambda}^{1k}\partial_r\omega_k = 0$,
{\it i.e.}, the unphysical Taub-NUT charge $a$ is zero, relates the
two boost parameters $\delta_{1,2}$ in the following way:
\begin{equation}
|P|{\rm tanh} \delta_2 + |Q|{\rm tanh} \delta_1 = 0 .
\label{boostcon}
\end{equation}
Thereby, the transformed solutions (\ref{gensol}) are parameterized
by 4 independent parameters, {\it i.e.}, the non-extremality
parameter
\footnote{When the non-extremality parameter $\beta$ is zero and the
other parameters are kept finite, the no-Taub-NUT-charge condition
(\ref{boostcon}) ensures $\vec{\cal P} \cdot \vec{\cal Q} = 0$,
{\it i.e.}, this is a  condition satisfied by supersymmetric
configurations \cite{SUPER}.}
$\beta$, the electric $Q$ and magnetic $P$ charges of the
$U(1)_M\times U(1)_E$ solution, and the boost parameters
$\delta_{1,2}$, subject to the constraint (\ref{boostcon}).
The resultant solution is in turn specified by the mass
\footnote{When no-Taub-NUT-charge condition (\ref{boostcon}) is
imposed, the mass $M$ is compatible with the corresponding
Bogomol'nyi bound: $M \ge |\vec {\cal P}|+|\vec{\cal Q}|$.}
$M$ and four charges, however only three of them are independent.

The remaining $2n-3$ degrees of freedom, required to parameterize
the most general, static, spherically symmetric BH's in Abelian
$(4+n)$-d KK theory, are then  provided by  $SO(n)/SO(n-2)$ rotations
on the solutions (\ref{gensol}).

We shall now analyze the global space-time structure and the thermal
properties of the above solution.  Since $SO(n)/SO(n-2)$ rotations
on (\ref{gensol}) do not change the 4-d space-time (as well as
$\varphi$ and the scalar product $\vec{\cal P}\cdot \vec{\cal Q}$),
it is sufficient to consider the solutions (\ref{gensol}) for the
purpose of determining the space-time  and thermal properties for all
the $(4+n)$-d Abelian KK BH's.  Without loss of generality,
we assume that $|Q| \geq |P|$.  In the case of $|Q| \leq |P|$,
the roles of $(\delta_1,\delta_2)$ and $(P,Q)$ are interchanged.

We first discuss the singularity structure.  Non-extreme solutions
($\beta>0$) always have a space-time singularity behind or at
$r=-2\beta$.  Namely, the space-time singularity, {\it i.e.}, the
point at which $R(r)=0$, where the Ricci scalar $\cal{R}$ blows up,
occurs at the real roots of $X(r)$ and $Y(r)$, which are always
$\le -2\beta$ with equality holding when $P=0$ or $\delta_2=0$.
On the other hand, $\lambda(r)$ is zero at $r=0$ and $r=-2\beta$,
provided  $X(r)$ and $Y(r)$ do not have roots at these points,
{\it i.e.}, when $\delta_2 \neq 0$ and $P \neq 0$, in which case $r=0$
and $r=-2\beta$ correspond to the outer and inner horizons,
respectively.

The extreme limit ($\beta\rightarrow 0$) with the other parameters
finite corresponds to supersymmetric BH's with the singularity at $r=0$.
The singularity is null, {\it i.e.}, $r=0$ is also the horizon, except
when $P=0$, in which case the singularity becomes naked.  The extreme limit
($\beta\rightarrow 0$) with $|Q|\rightarrow |P|$, while keeping $\beta{\rm
e}^{2|\delta_2|}\equiv 2|q|$ and $||Q|-|P||{\rm e}^{2|\delta_2|}\equiv
4|\Delta|$ finite, corresponds to nonsupersymmetric BH's with
the global space-time of extreme Reissner-Nordstr\" om BH's.

Thermal properties of solutions (\ref{gensol}) are specified by
the 4-d space-time at the outer horizon located at $r=0$.
The Hawking temperature\cite{HAW} $T_H = |\partial_r\lambda(r=0)|/4\pi$
is given by
\begin{eqnarray}
T_H &=& {1 \over {4\pi\left(\hat{P}\hat{Q}\right)^{1/2}
{\rm cosh}\delta_1 {\rm cosh}\delta_2}}\nonumber\\
&=& {{[|Q|^2{\rm cosh}^2 \delta_2
- |P|^2{\rm sinh}^2 \delta_2]^{1/2}} \over
{4\pi\left(\hat{P}\hat{Q}\right)^{1/2}|Q|{\rm cosh}^2 \delta_2}}.
\label{temp}
\end{eqnarray}
As the boost parameter $\delta_2$ increases the temperature $T_H$
decreases, approaching zero temperature.  In the supersymmetric extreme
limit and with zero $P$, the temperature is always infinite
independently of $\delta_2$.  In the non-supersymmetric extreme limit,
the temperature is zero.

The entropy\cite{BEK} $S$ of the system, determined as $S = {1\over 4}
\times$(the surface area of the event horizon), is of the following
form:
\begin{eqnarray}
S &=& 2\pi \beta \left(\hat{P}\hat{Q}\right)^{1/2}{\rm cosh}\delta_1
{\rm cosh}\delta_2
\nonumber\\
&=& {{2\pi\beta \left(\hat{P}\hat{Q}\right)^{1/2}|Q|{\rm cosh}^2
\delta_2}
\over {[|Q|^2{\rm cosh}^2 \delta_2 -|P|^2{\rm sinh}^2
\delta_2]^{1/2}}} .
\label{entro}
\end{eqnarray}
The entropy increases with $\delta_2$, approaching infinity [finite
value] as $\delta_2 \to \infty$ [non-supersymmetric extreme limit
is reached].  In the supersymmetric extreme limit, the entropy is zero.

We now summarize the results according to the values of parameters
$\delta_2$, $P$ and $\beta$:
\begin{itemize}
\item Non-extreme BH's with $\delta_2 \ne 0$, $P \ne 0$
\footnote{Note, that non-extreme BH's of 5-d KK theory belong
to this class.  They are obtained from solutions (\ref{gensol})
by performing an $SO(2)$ rotation on the $(n+1)$-$th$ and $(n+2)$-$th$
indices of the corresponding matrix $\chi$, however, the corresponding
rotation parameter is related to $\delta_2$.}:
The global space-time is that of non-extreme Reissner-Nordstr\" om
BH's, {\it i.e.}, the time-like singularity is hidden behind the inner
horizon.  The temperature $T_H$  [entropy $S$] is finite, and decreases
[increases] as $\delta_2$ or  $\beta$ increases,
approaching zero temperature [infinite entropy].
\item Non-extreme BH's with $\delta_2 = 0$ or $P = 0$:
The singularity structure is that of the Schwarzschild BH's, {\it i.e.},
the space-like singularity is  hidden behind the (outer) horizon.
The tempterature $T_H$ [entropy $S$] is finite and decreases
[increases] as $\beta$ increases, approaching zero [infinity].
\item  Supersymmetric extreme BH's, {\it i.e.}, $\delta_2$ finite:
For $P\ne 0$, the solution has a null singularity, which becomes
naked  when $P=0$.  The temperature $T_H$ [entropy $S$] is finite
and becomes infinite [zero] when $P=0$.
\item  Non-supersymmetric extreme BH's , {\it i.e.},
$|\delta_2| \to \infty$ with ($q$,$\Delta$) non-zero:latex
The global space-time is that of extreme Reissner-Nordstr\" om BH's
with zero temperature $T_H$ and finite entropy S
\footnote{Extreme dyonic solutions of 5-d KK theory \cite{GW} are
obtained from this one by choosing an $SO(2)$ rotation angle, related
to $Q$, $q$ and $\Delta$.}.
\end{itemize}

\acknowledgments
The work is supported by U.S. DOE Grant No. DOE-EY-76-02-3071,
the NATO collaborative research grant CGR 940870, and NSF Career
Advancement Award  PHY95-12732.

\end{document}